\documentclass[10pt,a4paper,twocolumn]{article}
\usepackage{mathptmx} 
\usepackage{graphicx} 
\usepackage{amstext} 
\usepackage{amsmath}
\usepackage[top=25mm,bottom=37mm,left=20mm,right=20mm,columnsep=10mm]{geometry} 
\usepackage{color} 
\definecolor{myblu}{rgb}{0.1,0.1,0.5}
\usepackage{hyperref} 
\usepackage{sectsty,textcase} 
\sectionfont{\large\MakeTextUppercase} 
\usepackage{secdot} 
\usepackage{textcomp}
\usepackage{upgreek}

\hypersetup{
    unicode=false,          
    pdftoolbar=true,        
    pdfmenubar=true,        
    pdffitwindow=false,     
    pdftitle={Design of a Freely Rotating Wind Tunnel Test Bench for Dynamic Coefficients Measurements},    
    pdfauthor={Dr. Muller Laurène and Dr. Libsig Michel},     
    pdfsubject={Aerodynamics},   
    pdfcreator={},   
    pdfproducer={},  
    pdfkeywords={Aerodynamics,} {Flight Mechanics,} {3 DoF,} {Rigid Body Dynamics,} {RBD,} {Numerical Simulation,} {CFD,} 
                {Stereovision,} {Supersonic,} {Wind Tunnel,} {Free-Flight,} {Shock Tunnel,} {Ludwieg Tube,} {DREV-ISL,}
                {Projectile,} {Image Processing,} {Dynamic Aerodynamic Coefficients} , 
    pdfnewwindow=true,      
    colorlinks=true,       
    linkcolor=black,          
    citecolor=black,        
    filecolor=magenta,      
    urlcolor=blue           
}

\begin{document}
\hbadness=10000
\vbadness=10000
\title{\vspace{-18mm}
\begin{minipage}{\linewidth}
\hspace{5mm}\raisebox{-50pt}{\includegraphics[width=.23\textwidth]{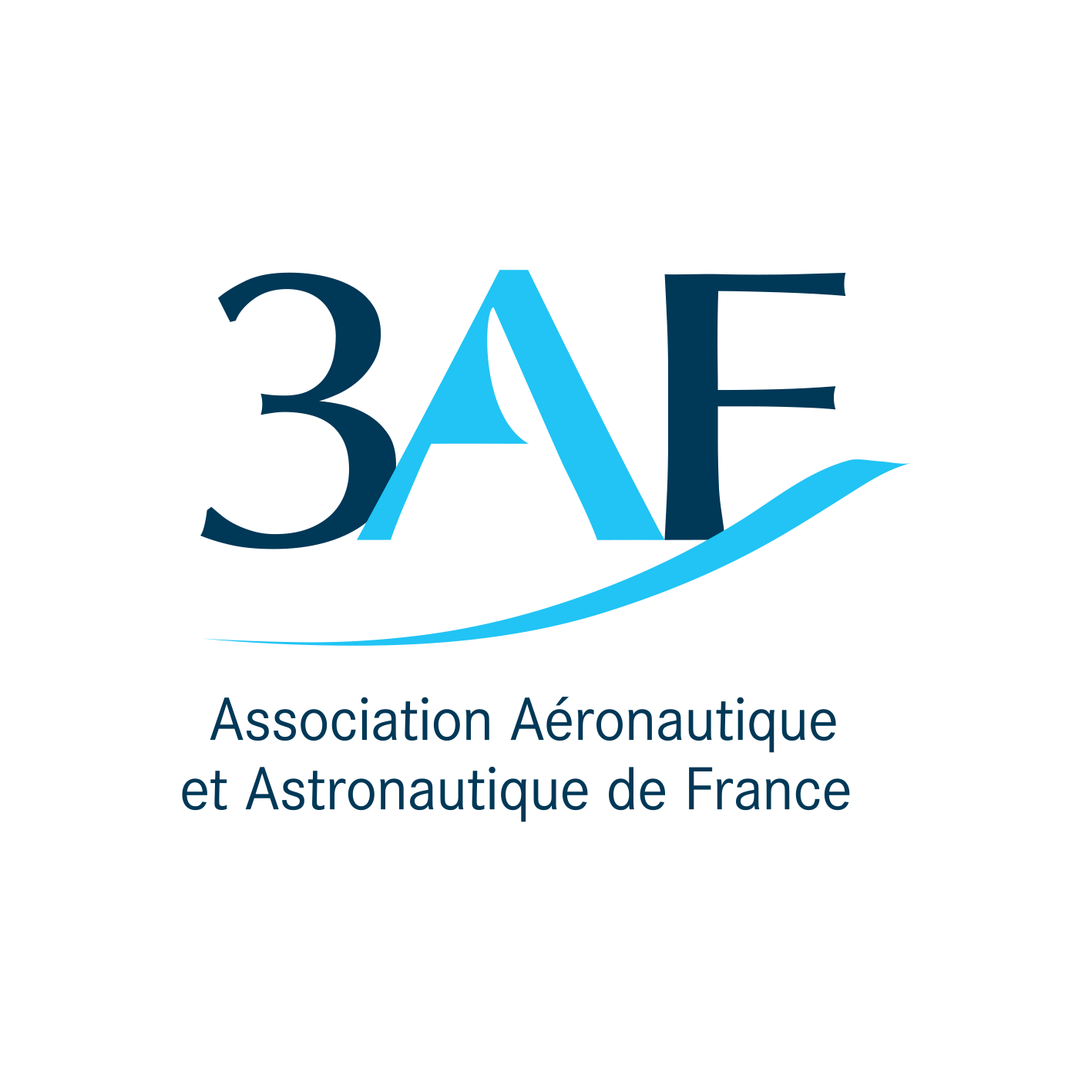}}\hspace{4mm}
\textcolor{myblu}{\textbf{\textit{\normalsize\begin{tabular}{l}
56$^\text{th}$ 3AF International Conference\\ on Applied Aerodynamics\\ 28 --- 30 March 2022, Toulouse -- France
\end{tabular}}}}
\hspace{16mm}\textbf{\normalsize FP48-AERO2022-Muller}
\end{minipage}\\\vspace{10mm}
\textbf{\Large DESIGN OF A FREELY ROTATING WIND TUNNEL TEST BENCH FOR MEASUREMENTS OF DYNAMIC COEFFICIENTS}}
\author{\textbf{\normalsize Muller Laur{\`e}ne$^\text{(1)}$ and Libsig Michel$^\text{(2)}$}
\\{\normalsize\itshape
$^\text{(1)}$ISL, 5 rue du G{\'e}n{\'e}ral Cassagnou 68300 Saint-Louis (France), laurene.muller@isl.eu}
\\{\normalsize\itshape
$^\text{(2)}$ISL, 5 rue du G{\'e}n{\'e}ral Cassagnou 68300 Saint-Louis (France), michel.libsig@isl.eu}}
\date{}

\newcommand{\Cma}{$C_{m\alpha}$ \,}
\newcommand{\CmqCma}{$C_{mq}+C_{m\dot{\alpha}}$ \,}
\newcommand{\CmaNoSpace}{$C_{m\alpha}$}
\newcommand{\CmqCmaNoSpace}{$C_{mq}+C_{m\dot{\alpha}}$}
\newcommand{\graphicsvspace}{\vspace{-0.2cm}}
\newcommand{\captionvspace}{\vspace{-0.7cm}}
\newcommand{\figvspace}{\vspace{-0.3cm}}
\newcommand{\eqvspace}{\vspace{-0.2cm}}
\newcommand{\itemizevspace}{\vspace{-0.25cm}}
\newcommand{\tabcaptionvspace}{\vspace{-0.3cm}}

\maketitle

\hbadness=5000
\vbadness=5000

\begin{abstract}
The needs to improve performances of artillery projectiles require accurate aerodynamic investigation methods. The aerodynamic design of a projectile usually starts from numerical analyses, mostly including semi-empirical methods and/or Computational Fluid Dynamics (CFD), up to experimental techniques composed of wind-tunnel measurements or free-flight validations. In this framework, the present paper proposes a dedicated measurement methodology able to simultaneously determine the stability derivative \Cma and the pitch damping coefficient sum \CmqCma in a wind tunnel by means of a single and almost non-intrusive metrological setup called MiRo. This method is based on the stereovision principle and a three-axis freely-rotating mechanical test bench. In order to assess the reliability, repeatability and accuracy of this technique, the MiRo wind tunnel measurements are compared to other sources like aerodynamic balance measurements, alternative wind tunnel measurements, Ludwieg tube measurements, free-flight measurements and CFD simulations.
\end{abstract}

\section{INTRODUCTION}

ISL (French-German Research Institute of Saint-Louis) initiated the development of a three-axis freely rotating test bench for projectiles for the validation of concepts and pitch damping coefficient measurements in wind tunnels. The final goal of this test bench, called MiRo, is to investigate the attitude of spin-stabilized models containing decoupled actuators \cite{Martinez2015}. Due to the mechanical complexity of such a technology, the design of MiRo is performed step by step, thanks to CFD simulations and supersonic blow-down wind-tunnel experiments. The measurement of the projectile's motion during the blow-down is performed by a stereovision technique, based on two high-speed cameras. Recordings from both cameras can be processed frame by frame and coupled thereafter by means of an image processing code \cite{Muller2019}, resulting in the evolution of the attitude of the model as a function of the time. The stability derivative and the pitch damping moment coefficient are identified from this signal thanks to a curve fitting algorithm, based on a mathematical motion model. The effects of the holding mechanism on the MiRo measurements obtained during a previous test campaign has already been evaluated in \cite{Muller2020}.

\begin{figure}[h]
	\centering
	\graphicsvspace
	\includegraphics[width=\linewidth]{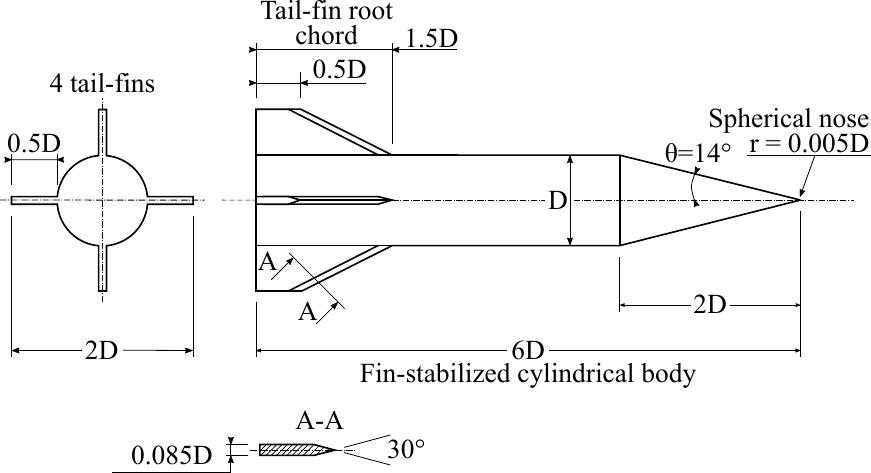}
	\captionvspace
	\caption{The DREV-ISL model}
	\label{fig:DREV_ISL_model}
	\figvspace
\end{figure}

The study presented in this paper consists in validating the feasibility of using a freely-rotating measurement technique and assessing the resulting uncertainty. This first design step has exclusively been performed on a statically stable ammunition on which the roll and yaw motions have been locked. Hence, the work has been carried out on a single degree of freedom (1DoF). This validation consists in confronting the MiRo measurements with other commonly used experimental and numerical techniques, such as:
\itemizevspace
\begin{itemize}
\itemsep-.3em 
\item[-] wind tunnel and Ludwieg tube measurements using wire-suspended 1DoF freely rotating model,
\item[-] free-flight measurements based on an optical stereovision technique,
\item[-] unsteady flow simulations in which the model is submitted to a forced oscillating motion,
\item[-] coupling of rigid body dynamics (RBD) and unsteady CFD simulations in order to investigate a freely oscillating virtual model.
\end{itemize}
\itemizevspace
This paper is organized in three parts. The first part consists in describing the MiRo experimental setup, distinguishing the mechanical test bench, the optical setup and the post-treatment methodology. The experimental facilities and measurement techniques enumerated above as well as the CFD simulations used for the comparison are described in the second part. And finally, the results will be confronted to each other and analysed in the last part.
Due to a large amount of already existing data, the fin-stabilized DREV-ISL rocket \cite{Dupuis1993}\cite{Berner1993} schematized in Fig. \ref{fig:DREV_ISL_model} has been retained. For mechanical and experimental reasons, the calibre of the DREV-ISL model has been set to 40 mm.

\section{DESCRIPTION OF THE MIRO MEASUREMENT TECHNIQUE}

The purpose of the MiRo technique is to measure three-dimensional static- and dynamic aerodynamic coefficients in the wind tunnel. Therefore, the 3D attitude of a freely rotating model has to be captured and the resulting signal has to be post processed in order to identify the coefficients. Thus, MiRo is composed of three main blocks:
\itemizevspace
\begin{itemize}
\itemsep-.3em 
\item[-] a dedicated wind tunnel mechanical test bench,
\item[-] a stereovision-based attitude measurement coupled with an image processing algorithm,
\item[-] an aerodynamic coefficient identification algorithm.
\end{itemize}
\itemizevspace

\subsection{Wind tunnel test bench}

\begin{figure}[t]
	\centering
	\includegraphics[width=\linewidth]{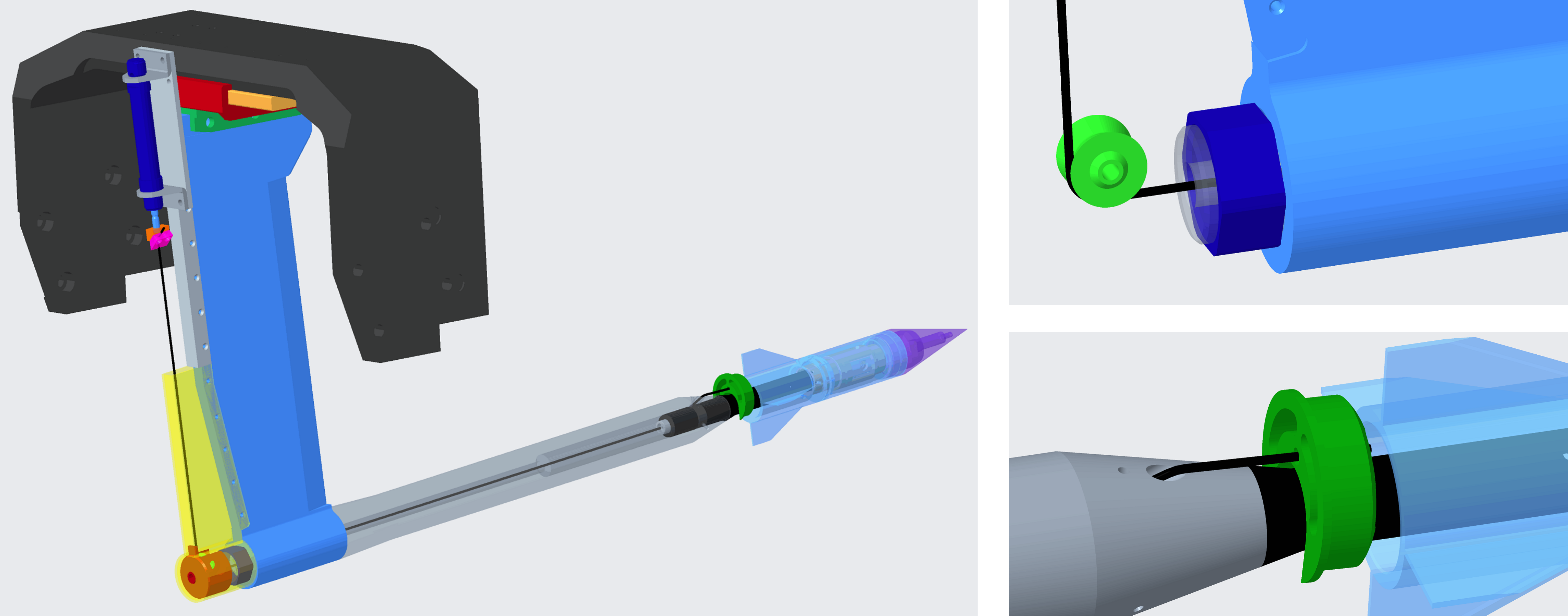}
	\captionvspace
	\caption{The MiRo wind tunnel test bench}
	\label{fig:MiRoWTBenchFull}
\end{figure}

\begin{figure}[t]
	\centering
	\includegraphics[width=\linewidth]{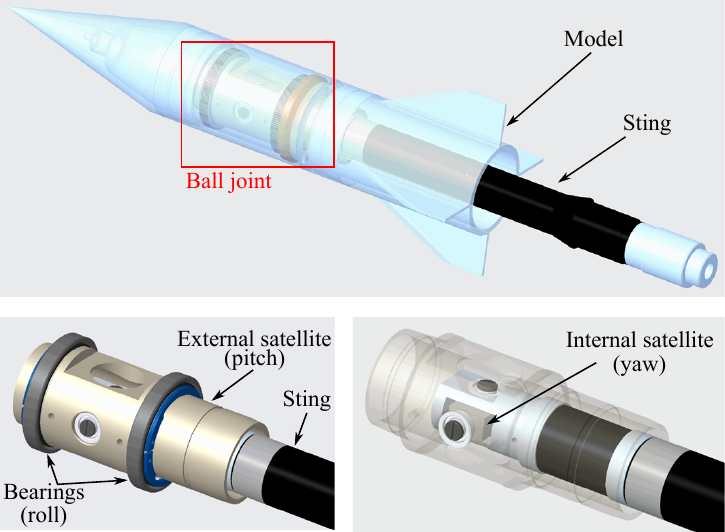}
	\captionvspace
	\caption{The MiRo holding mechanism}
	\label{fig:MechanicalTestBenchFull}
\end{figure}

As shown in Figs. \ref{fig:MiRoWTBenchFull} and \ref{fig:MechanicalTestBenchFull}, the model of the projectile is held in the stream by means of a sting and a ball joint located at its centre of gravity. The model must be hollowed out, starting from the base, in order to mount the ball joint inside. In the 3DoF version of MiRo, the ball joint is composed of two bearings and two freely rotating satellites:
\itemizevspace
\begin{itemize}
\itemsep-.3em 
\item[-] The internal satellite is mounted on the sting and allow rotations around the yawing axis,
\item[-] The external satellite is mounted on the internal satellite and allows rotations around the pitching axis,
\item[-] Both bearings are mounted on the external satellite and allow rotations along the rolling axis.
\end{itemize}
\itemizevspace
The angular amplitudes of the yawing and pitching rotations are of about 2$^{\circ}$, and the model is balanced by means of additional weights.

The aerodynamic coefficients can only be measured if the projectile is subjected to a damped oscillating attitude. Therefore, a specific system has been designed in order to maintain the projectile at a non-zero initial angle of attack during the start-up of the wind tunnel. Once the flow has stabilized, the model is released in order to oscillate freely around its three axes. The incidence of the model is held by the dark green 3D printed plastic part which is linked to the dark blue pneumatic cylinder thanks to a wire (Fig. \ref{fig:MiRoWTBenchFull}). When the model has to be released, the cylinder pulls the wire thanks to a user-triggered solenoid valve. The light green pulley ensures that no friction occurs at the corner of the mounting structure.

\subsection{Measurement of the 3-dimensional motion}

\begin{figure}[h]
	\graphicsvspace
	\centering
	\includegraphics[width=\linewidth]{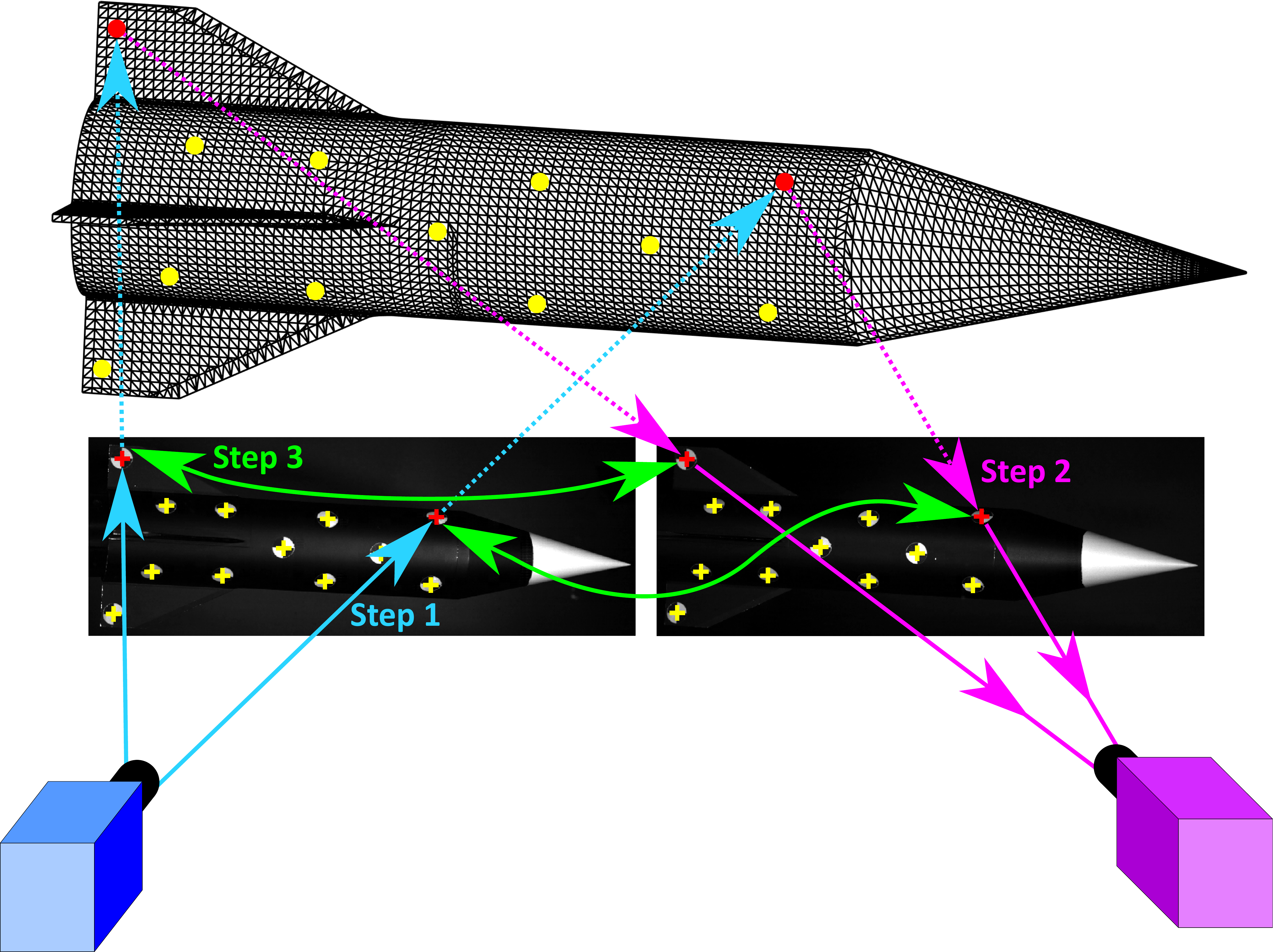}
	\captionvspace
	\caption{The stereovision and image processing principles}
	\label{fig:StereovisionPrinciple}
	\figvspace
\end{figure}

In order to determine the motion of the model in the three directions of the physical space, the stereovision principle is employed. This optical technique is designed to capture the depth of the scene, like the human brain does naturally by combining information sent from both eyes. As illustrated in Fig. \ref{fig:StereovisionPrinciple}, this goal can be reached in observing markers on the investigated object from different locations with at least two cameras. The resulting recordings are processed in order to evaluate the depth using a numerical process. Therefore a new frame of reference called the pixel coordinate system has to be introduced. It corresponds to the reference frame of the picture being displayed on the computer and refers to the coordinates of the pixels by means of two components: the line number and the column number \cite{Liu2000}.

To obtain an accurate attitude reconstruction, Secchi markers are placed on the model in order to be tracked using a corner detection algorithm \cite{Harris1988}. The exact positions of the marker's centres in the pictures are determined and these markers are coupled frame by frame. Knowing their position in the pixel reference frame, the 3D scatterplot corresponding to the 3D position of the markers can be evaluated by modelling the cameras with the pinhole model.

The pinhole camera is a black box which contains an aperture like a small hole. It reproduces an image after the passage of the light beams through the orifice. On the other hand, the pinhole camera model corresponds to the mathematical formulation which links the coordinates of physical points in the 3D space to their projections in the pixel coordinate system. Therefore, the intersections of 3D optical lines (light beams) emanating from the cameras have to be determined with an optimization algorithm. Geometric distortions are not considered in this model.

\begin{figure}[t]
	\centering
	\includegraphics[width=\linewidth]{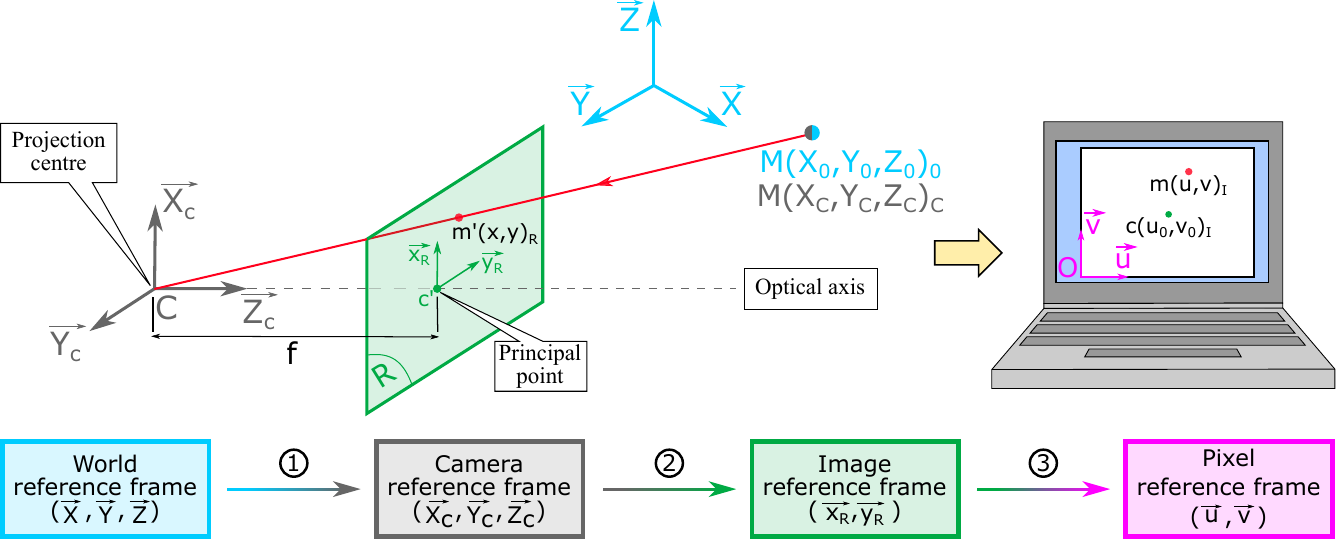}
	\captionvspace
	\caption{Decomposition of the pinhole model}
	\label{fig:Pinhole_Decomposition}
\end{figure}

Fig. \ref{fig:Pinhole_Decomposition} shows the decomposition of the pinhole model. It consists in performing three successive elementary transformations linking the coordinates of the physical 3D point $M$ expressed in the world reference frame $(X, Y, Z)$ to the pixel coordinates $(u, v)$ of the point $m$ on the screen. Eq. \ref{eq:PinholeModel} provides the resulting mathematical expression \cite{Muller2019} of the pinhole model applied to a single camera. 
\begin{equation}
	s
	\begin{pmatrix}
		u \\ v \\ 1
	\end{pmatrix}
	= 
	\begin{bmatrix}
		\alpha_u & 0 & u_0 & 0 \\ 0 & \alpha_v & v_0 & 0 \\ 0 & 0 & 1 & 0
	\end{bmatrix}
	\begin{bmatrix}
		R_{3\times3} & T_{3\times1} \\ 0_{1\times3} & 1
	\end{bmatrix}
	\begin{pmatrix}
		X \\ Y \\ Z \\ 1
	\end{pmatrix}
	\label{eq:PinholeModel}
\end{equation}

\begin{figure}[h]
	\graphicsvspace
	\centering
	\includegraphics[width=\linewidth]{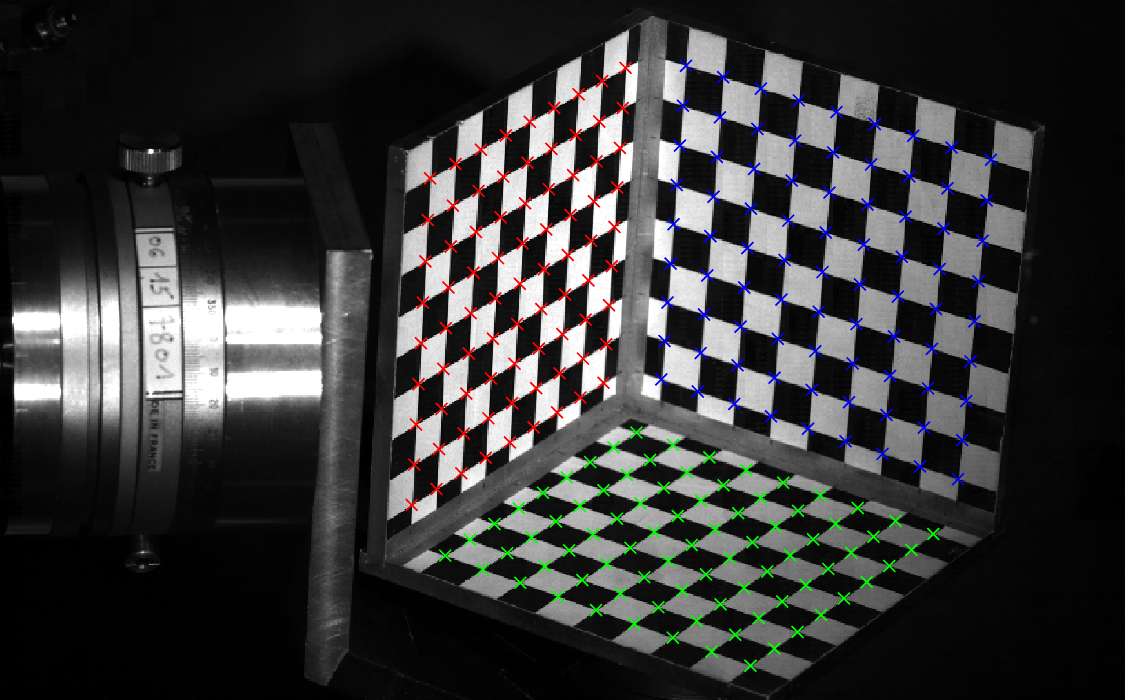}
	\captionvspace
	\caption{Detection of the checkerboard's control points in the $X = 0$ (red), $Y = 0$ (green) and $Z = 0$ (blue) planes}
	\label{fig:Detection_coin_mires}
	\figvspace
\end{figure}

The so called intrinsic parameters $\alpha_u$, $\alpha_v$, $u_0$ and $v_0$ are specific to the lens of the camera, while the three Euler angles of the rotation matrix $R_{3\times3}$ and the three components of the translation vector $T_{3\times1}$ are the extrinsic parameters that express the camera position with respect to the object. These ten parameters are determined using the calibration process of Heikkila and Silven \cite{Heikkila1997} with the 3D raw card of Fig. \ref{fig:Detection_coin_mires}. This raw card is composed of three faces covered with checkerboard patterns. The 3D positions of the corners of the checkerboards are known insofar as the size of the squares and the angular orientations of the planes are known.

Each marker creates an image point on both recordings. The pixel coordinates of the centre of each single marker on camera \#1 and \#2 are noted $U_1 = (u_1, v_1, 1)$ and $U_2 = (u_2, v_2, 1)$. Hence, the stereoscopic Eq. \ref{eq:PinholeModel} can be written as a function of the physical coordinates $[X] = (X, Y, Z)$ in a more compact form:
\begin{equation}
	\left\{
	\begin{array}{c}
	s_1 U_1 = I_{C1} \left(  R_1 \cdot \left[ X \right] + T_1 \right) \\
	s_2 U_2 = I_{C2} \left(  R_2 \cdot \left[ X \right] + T_2 \right) 
	\end{array}
	\right.
	\label{eq:StereoSystem}
\end{equation}
In this case $s_i$ is the scale factor, $I_{Ci}$ is the intrinsic parameters matrix, $R_i$ is the rotation matrix and $T_i$ is the translation vector between the world- and camera reference frames. Excepting $s_i$, all these parameters are determined during the calibration process \cite{Heikkila1997}.

Both relations of the system of equations \ref{eq:StereoSystem} correspond to matrix equations of the light beams (3D lines) in space. As Fig. \ref{fig:StereovisionPrinciple} illustrates, solving the coordinates $\left[ X \right]$ of the centre of the marker consists in determining the intersection of the respective lines emanating from both cameras. This system of equations is overdetermined because it contains six scalar equations and five unknown factors: the three scalar values of $\left[ X \right]$, $s_1$ and $s_2$. The unknown variables can be isolated by rearranging the system in a matrix form, such as:
\begin{equation}
	\underbrace{
	\begin{bmatrix}
	 R_1 & -I_{C1}^{-1}U_1 & 0_{3\times1} \\
	 R_2 & 0_{3\times1}    & -I_{C2}^{-1}U_2
	\end{bmatrix}
	}_P
	\cdot
	\begin{bmatrix}
	 \left[ X \right] \\ s_1 \\ s_2
	\end{bmatrix}
	=
	\underbrace{-
	\begin{bmatrix}
	 T_1 \\ T_2
	\end{bmatrix}
	}_P
	\label{eq:StereoSystemMatrix}
\end{equation}

With $P$ being a 6-component vector, and $H$ a 6$\times$5 matrix

The least square matrix solution of Eq. \ref{eq:StereoSystemMatrix} can be expressed as \cite{levenberg1944method}\cite{marquardt1963algorithm}:
\begin{equation}
	\begin{bmatrix}
	 \left[ X \right] \\ s_1 \\ s_2
	\end{bmatrix}
	= - \left( H^T \cdot H \right)^{-1} \cdot H^T \cdot P
	\label{eq:StereoSystemSolution}
\end{equation}

\subsection{Determination of the static and dynamic aerodynamic coefficients}
\label{part:DeterminationCoefficients}

\begin{figure}[h]
	\graphicsvspace
	\centering
	\includegraphics[width=\linewidth]{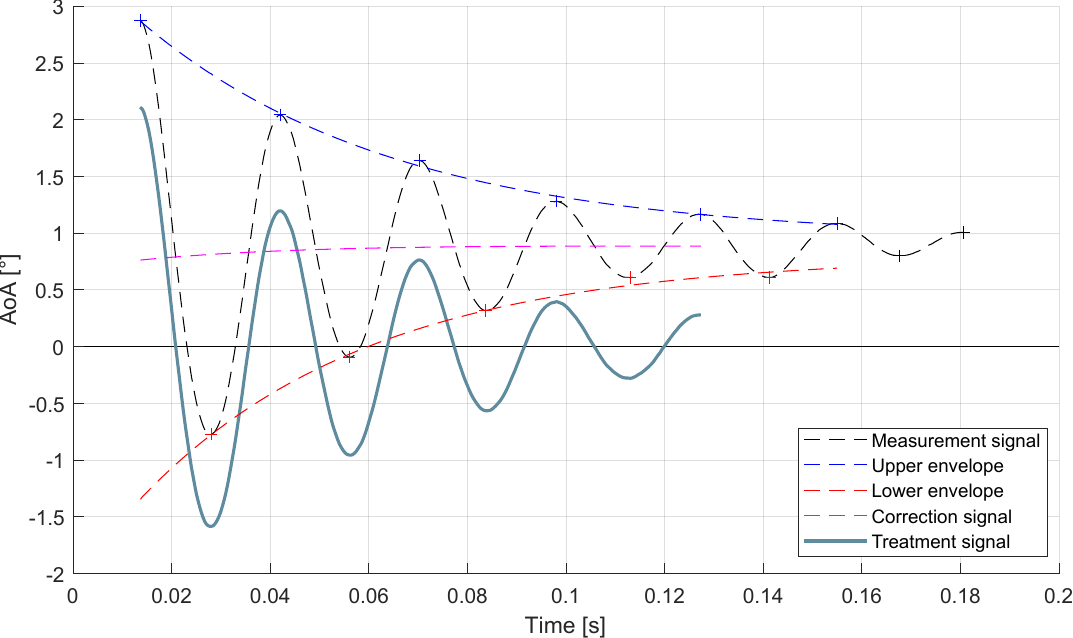}
	\captionvspace
	\caption{Correction of the measurement signal}
	\label{fig:AoA_Correction_Method}
	\figvspace
\end{figure}

In order to determine the stability derivative \Cma and the pitch damping coefficient sum \CmqCmaNoSpace, the angle of attack of the stable DREV-ISL projectile follows a damped oscillating attitude during the blow down. For small amplitudes, constant flow velocities, negligible spin rates, the angular motion of the projectile can be described by the linearized angle of attack equation (Eq. \ref{eq:ModeleAoA}), as given by McCoy in \cite{McCoy1999}. Hence, the post processing algorithm for identifying aerodynamic coefficients analyses the evolution of the amplitude (Eq. \ref{eq:ModeleAoA_A}) and the frequency of the oscillations (Eq. \ref{eq:ModeleAoA_B}) in order to determine the \CmqCma and \Cma coefficients, respectively.
\begin{equation}
	\alpha \left( t \right) = \alpha_{max} \text{e}^{A t} \sin \left( B t + \phi_0 \right)
	\label{eq:ModeleAoA}
\end{equation}
with
\begin{equation}
	A = \frac{\rho S V_\infty D^2}{8 I_y} \left( C_{mq}+C_{m\dot{\alpha}} \right)
	\label{eq:ModeleAoA_A}
\end{equation}
\begin{equation}
	B = 2 \pi f = \sqrt{-\frac{\rho S V_\infty^2 D}{2 I_y} C_{m \alpha}}
	\label{eq:ModeleAoA_B}
\end{equation}
$I_y$ is the transverse inertia and $\alpha (t)$ the angle of attack in the plane of incidence. In order to estimate the aerodynamic coefficients, the damped sine wave curve (\ref{eq:ModeleAoA}) is superimposed on the measurement with the curve fitting algorithm. The estimation of the aerodynamic coefficients is performed by the identification of the fitting parameters $\alpha_{max}$, $\phi_0$, \Cma and \CmqCmaNoSpace.
The frame of reference in which the attitude measurement shall be performed is defined during the stereovision calibration process. Since this step is performed manually, a misalignment between the measurement- and aerodynamic reference frames may remain. For this reason the measured angle of attack curve must be corrected as illustrated in Fig. \ref{fig:AoA_Correction_Method}. Knowing that the balance angle of attack of the model is equal to $0^{\circ}$, an algorithm is used in order to center the movement around the $\alpha = 0^{\circ}$ axis (grey curve of Fig. \ref{fig:AoA_Correction_Method}). First, as the amplitude of the mathematical model is based on an exponential decay, two exponential function-based envelopes, in blue and red, are fitted on the upper and lower measurement peaks. Then, the pink correction signal, corresponding to the mean of both envelopes, is calculated. The grey corrected signal, on which the curve fitting algorithm is applied, is obtained by subtracting the pink correction signal from the black angle of attack curve obtained with the stereovision technique.

\section{DESCRIPTION OF THE COMPARISON METHODOLOGIES}
\label{part:DescriptionFull}

The MiRo measurements have been compared to experiments and CFD calculations as follows:
\itemizevspace
\begin{itemize}
\itemsep-.3em 
\item[-] wind tunnel measurements using a 1DoF freely rotating model hold by a wire from Mach 2.0 to 4.0,
\item[-] Ludwieg tube measurements using a 1DoF freely rotating model hold by a wire for Mach 3.0 and 4.5,
\item[-] free flight experiments for an initial velocity of Mach 2.0,
\item[-] forced oscillating motion using CFD from Mach 1.5 to 4.0,
\item[-] free oscillating motion using 1DoF/CFD coupling from Mach 1.5 to 4.0.
\end{itemize}
\itemizevspace

\subsection{Experimental facilities}

\subsubsection{Trisonic blow down wind tunnel}
\label{part:wind_tunnel}

\begin{figure}[h]
	\graphicsvspace
	\centering
	\includegraphics[width=\linewidth]{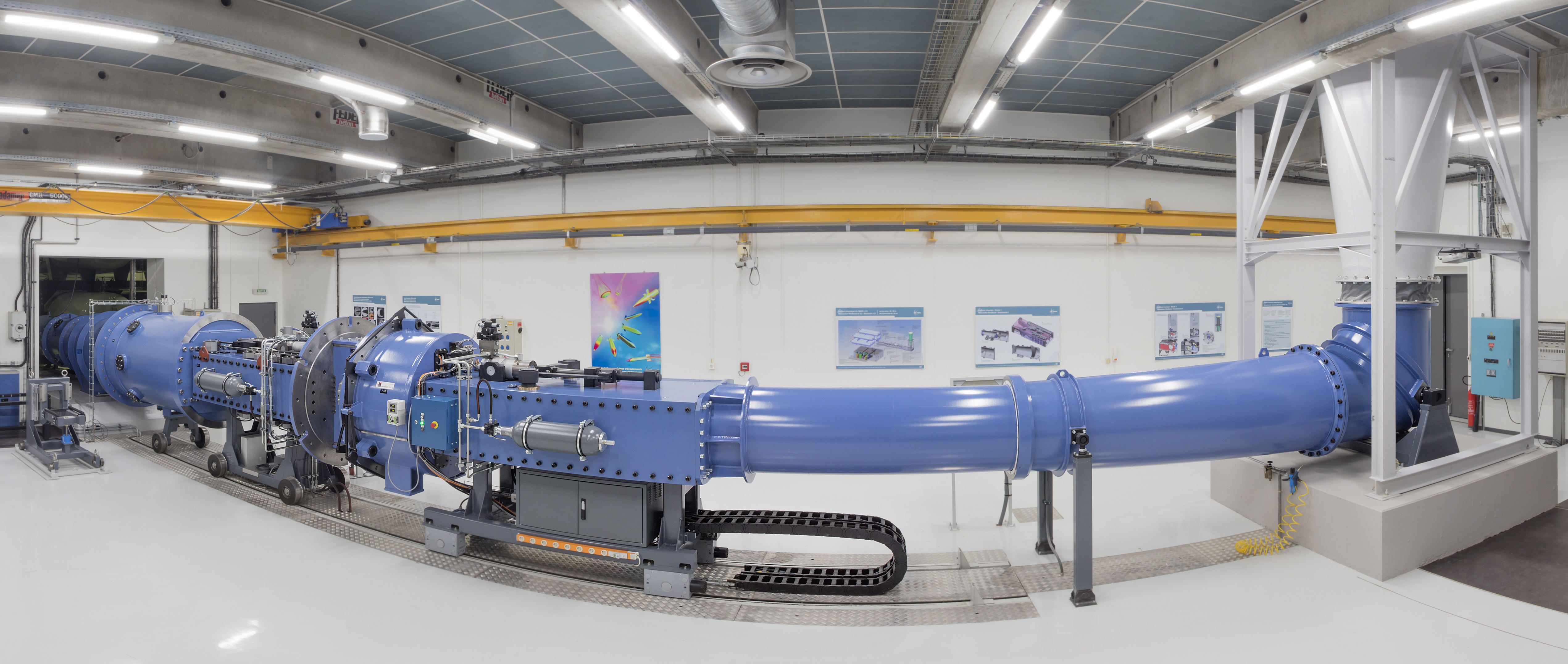}
	\captionvspace
	\caption{Fisheye picture of the trisonic wind tunnel}
	\label{fig:PhotoSoufflerie}
	\figvspace
\end{figure}

Fig. \ref{fig:PhotoSoufflerie} shows a picture of the ISL's trisonic blow down wind tunnel \cite{Delery2017}. This facility allows performing investigations for Mach numbers ranging from 0.5 to 4.0.

Air is compressed with external compressors up to a pressure of 30 bars in 288 m$^3$ reservoirs. During the blow down, the flow expands through a fast acting control valve and the user-defined total pressure is controlled. A tranquilization chamber destroys the turbulence vortices and the awaited Mach number is obtained thanks to a convergent-divergent Laval nozzle. The test section is located downstream of the nozzle exit, just after the grey flange on the left part of Fig. \ref{fig:PhotoSoufflerie}. Horizontal and vertical windows provide optical access all around the test section and dedicated apertures allow equipping the experiment with measurement devices. Further downstream, the air decelerates through a diffuser and is released to the atmosphere thanks to the vertical chimney.

\subsubsection{Ludwieg tube}

\begin{figure}[h]
	\graphicsvspace
	\centering
	\includegraphics[width=\linewidth]{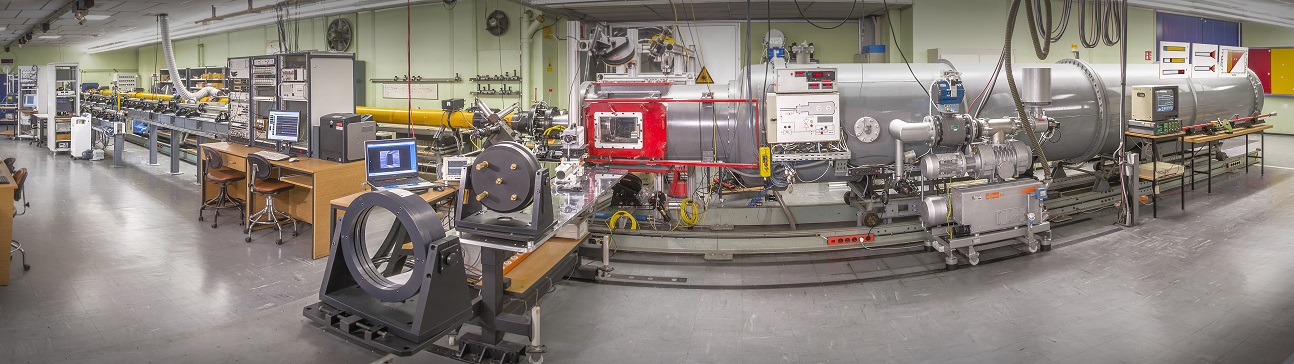}
	\captionvspace
	\caption{Fisheye picture of the shock tunnel}
	\label{fig:PhotoLudwiegTube}
	\figvspace
\end{figure}

Fig. \ref{fig:PhotoLudwiegTube} shows a picture of the ISL's shock tunnel facility \cite{Delery2017}. This facility allows performing investigations for Mach numbers ranging from 3.0 to 6.0. A second facility, not presented herein, extends the Mach number domain up to 11. For this experiment, the shock tunnel was modified in order to be driven as a Ludwieg tube \cite{ludwieg1969rohrwindkanal}. A diaphragm was inserted into the 18 meters long 100 mm yellow tube in order to separate the internal volume into two high- and low pressure zones. Upstream from the diaphragm, the air is compressed up to a calibrated disruption pressure. The disruption generates a subsonic flow and a shock wave propagating from the left to the right. A convergent-divergent Laval nozzle separates this part from the yellow tube with the test section in which the experiment is conducted. The desired Mach number is obtained by choosing the adequate nozzle geometry and the experiment is triggered once the flow reaches the model. The test duration depends on the length of the tube and the sonic speed of the injected gas. The flow inside the test section is stable until the expansion wave, resulting from the pressure balancing inside the tube, reaches the Laval nozzle.

\subsubsection{Open range test facility}

\begin{figure}[h]
	\graphicsvspace
	\centering
	\includegraphics[width=\linewidth]{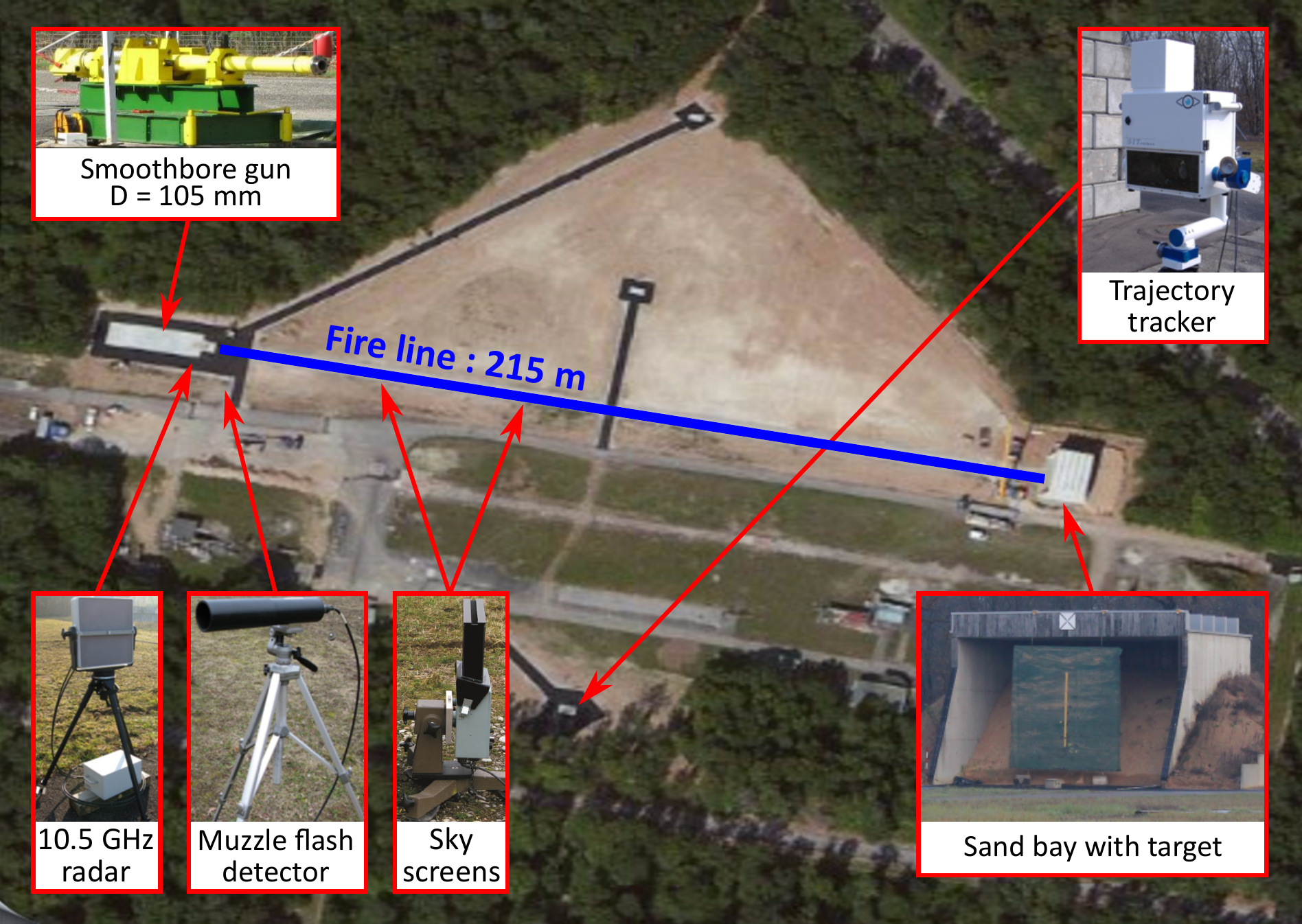}
	\captionvspace
	\caption{Bird's-eye view of the open range test facility}
	\label{fig:Champ_de_tir}
	\figvspace
\end{figure}

Fig. \ref{fig:Champ_de_tir} shows a bird's-eye view of the ISL's open range test facility \cite{Delery2017}. This facility allows shooting experiments to be conducted with 20 mm to 105 mm projectiles on ranges up to 1000 meters. Depending on the desired roll rate, smooth bore or rifled powder guns are employed and the projectiles can be launched with initial Mach numbers ranging from 0.6 to 5.0.

The experiments dedicated to this study have been carried out over a distance of 215 meters. In order to minimize the projectile's roll rate at the muzzle exit, the launches were performed with the 105 mm smoothbore gun. The flight of the projectile has been recorded using two trajectory trackers composed of high speed cameras and rotating mirrors dedicated to follow the projectile during the flight. In order to correct off-centred angular position of the mirrors, sky screens are installed along the fire line. The velocity of the projectile is measured via a 10.5 GHz radar and all the systems are triggered using a muzzle flash detector. At the end of the fire line, a sand bay aborts the flight of the projectile.

\subsection{Experimental test benches}

\subsubsection{MiRo experimental setup}

\begin{figure}[t]
	\centering
	\includegraphics[width=\linewidth]{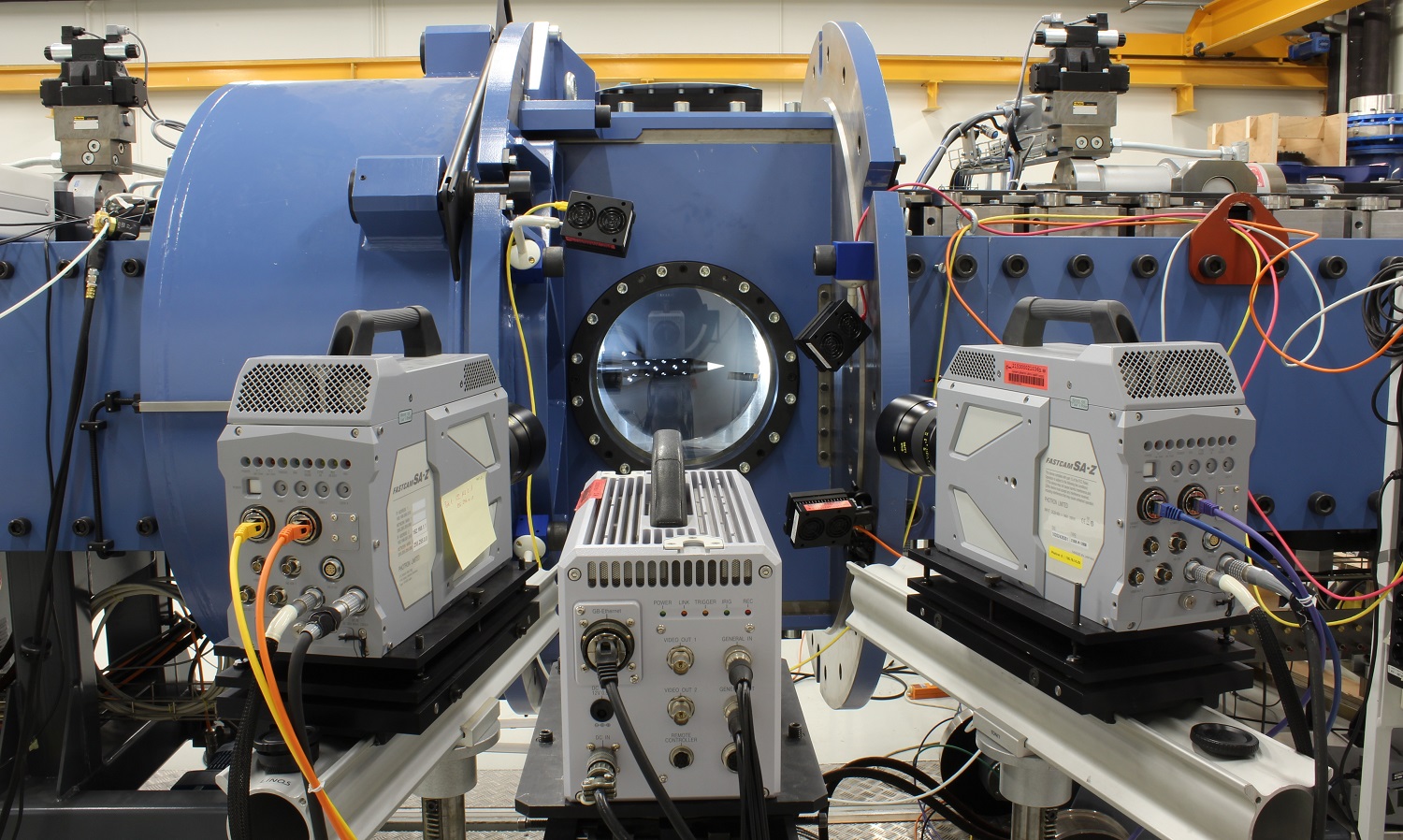}
	\captionvspace
	\caption{The MiRo experimental setup}
	\label{fig:Montage_MiRo_Reduced}
\end{figure}

Fig. \ref{fig:Montage_MiRo_Reduced} shows the MiRo test bench being mounted in the ISL's trisonic wind tunnel as described in part \ref{part:wind_tunnel}. A 4/2 solenoid valve used for the control of the pneumatic cylinder was connected on one side to a vacuum pump and on the other side to the compressed air distribution network. Using these pressure inputs, the cylinder was maintained in the open position during the start-up of the blow down. Once the model had to be released, a trigger box actuated the solenoid valve in order to pull the green holding part with the cylinder (Fig. \ref{fig:MiRoWTBenchFull}) and activated the high-speed cameras. 

\begin{figure}[h]
	\graphicsvspace
	\centering
	\includegraphics[width=\linewidth]{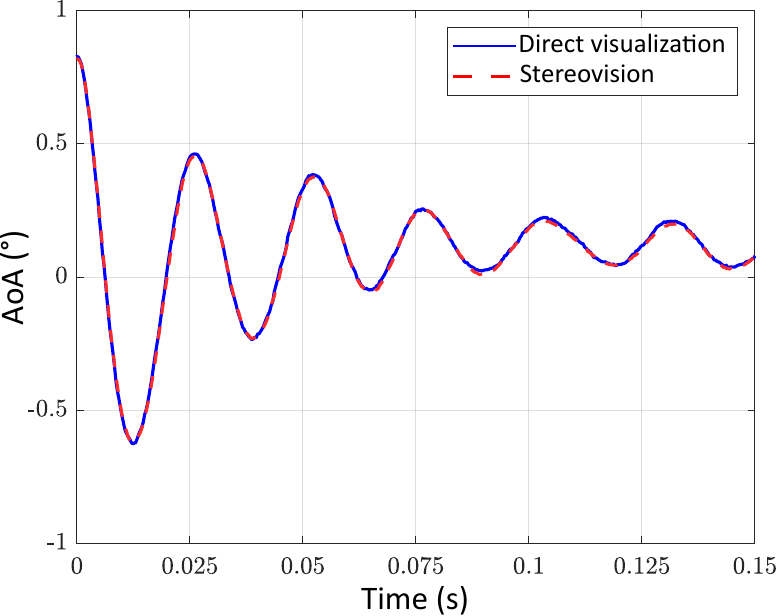}
	\captionvspace
	\caption{Angle of attack signal comparison using the stereovision and the direct visualization techniques}
	\label{fig:comparaison_stereo_visu}
	\figvspace
\end{figure}

The attitude of the projectiles was recorded using 2 Photron SA-Z high-speed cameras being able to record up to 20000 frames per second in a full frame format (1024$\times$1024 pixels) and an exposure time of 0.5 $\upmu$s. Four high-power GS Vitec MultiLED QT lamps have been used in order to illuminate the black model on which black and white Secchi markers were glued. Both cameras have been equipped with 105 mm lenses, placed at a distance of 1.2 meters from the model and spaced by 0.6 meter from each other. Using this configuration, the angle between both optical axes was equal to 30$^{\circ}$. An additional APX-RS Photron camera, in the middle of Fig. \ref{fig:Montage_MiRo_Reduced}, has been added between both SA-Z cameras in order to have a direct visualization for the validation of the pitching attitude obtained by the stereovision. Fig. \ref{fig:comparaison_stereo_visu} shows a comparison of angles of attack signals, using both techniques simultaneously, and shows that the pinhole-based stereovision algorithm provides a very good estimation of the projectile's attitude.

\subsubsection{Wire-suspended 1DoF freely rotating model}

\begin{figure}[ht]
	\graphicsvspace
	\centering
	\includegraphics[width=\linewidth]{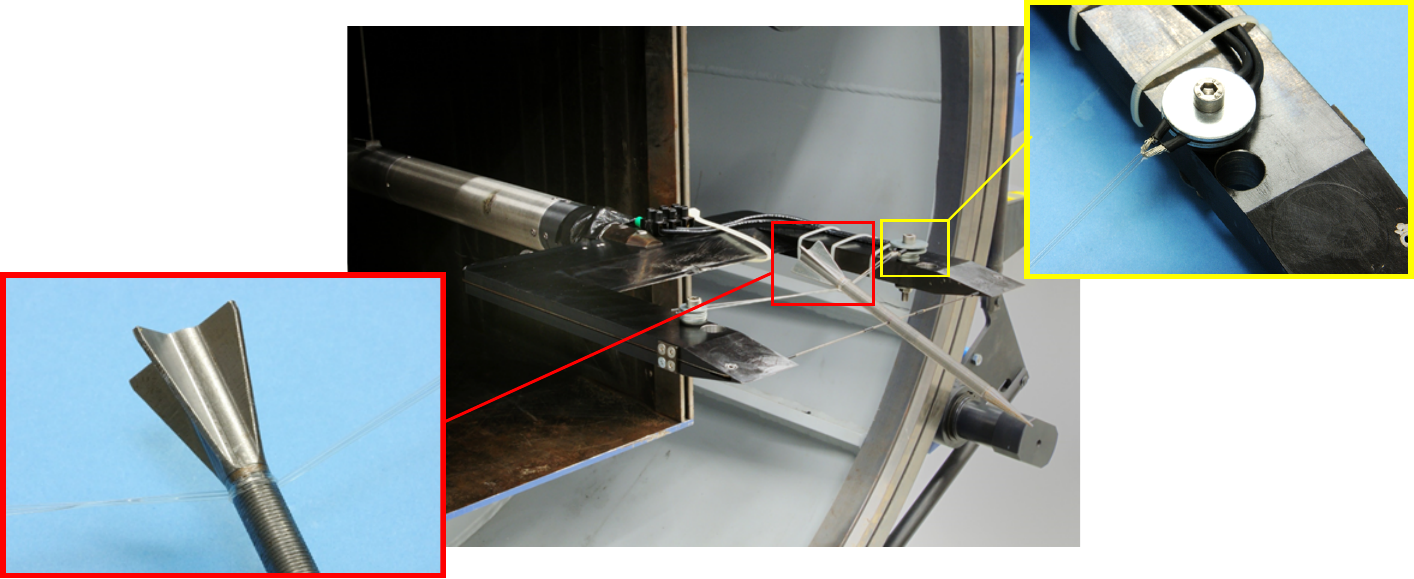}
	\captionvspace
	\caption{The wind tunnel test bench of the 1DoF freely rotating model held by a wire}
	\label{fig:Posseidon_2}
	\figvspace
\end{figure}

This measurement technique shows similarities with the MiRo technique because the projectile is kept at its centre of gravity and can rotate freely around its pitching axis. In spite of being a commonly used technique \cite{piper1960summary}\cite{shantz1960dynamic}, this mounting structure is much more intrusive because the holding wire interacts with the surrounding flow. Fig. \ref{fig:Posseidon_2} shows the wind tunnel test bench holding an armour piercing ammunition \cite{Michalski2021}, which was replaced by a DREV-ISL model for the present investigation. For the same reason as for the MiRo setup, the angle of attack has been maintained during the start-up and released once the flow has stabilized. This latter operation was achieved by means of nylon wire having been melt down by an electrical heating process.

\subsection{Numerical investigation methodologies}

The numerical simulations have been performed using the ANSYS Fluent CFD software. The external boundary conditions have been defined as pressure-far-fields and the non-slipping walls of the projectile have been considered to be adiabatic. The density and viscosity of the air have been obtained with the ideal gas law and the Sutherland equation, respectively. The calculations have been conducted in double precision and the turbulence has been modelled using the k$-\upomega$ \, SST turbulence model. Second order solvers have been used for all flow and turbulence variables.

Unstructured meshes were made with the Mosaic\texttrademark \, mesh generator of the Fluent Meshing software. The mosaic meshing technology consists in performing polyhedral connections between disparate mesh types. The volume has been filled with poly-hexacore cells (hexahedral cells in the fluid domain and poly-prism cells on the boundaries), whose sizes have been set using bodies of influence.

Three-dimensional compressible RANS simulations have systematically been performed in order to initialize the dynamic URANS simulations. The calculation of the \Cma and \CmqCma coefficients have been performed with the "forced oscillation motion" and the "1DoF/CFD coupling" techniques. 

\subsubsection{Forced oscillation motion}

\begin{figure}[t]
	\centering
	\includegraphics[width=\linewidth]{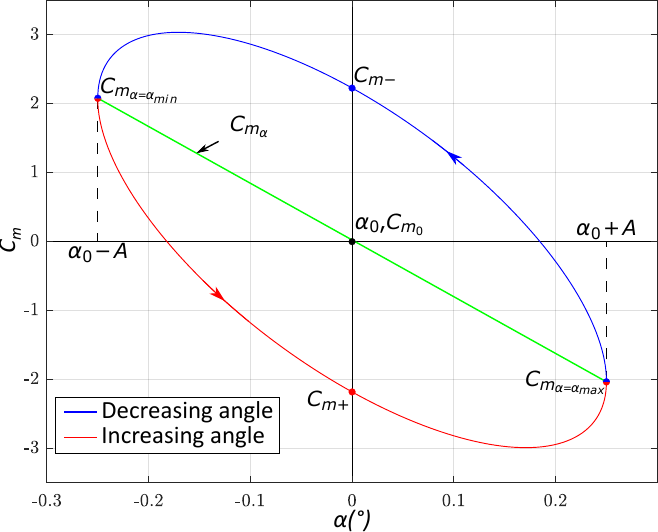}
	\captionvspace
	\caption{Pitching moment coefficient resulting from forced oscillation motion as a function of the angle of attack}
	\label{fig:ForcedOscillationCurve}
\end{figure}

The first methodology consists in forcing the pitching motion by using a low amplitude $\alpha (t) = \alpha_0 + A \cdot \sin ( \omega t )$ function while performing a URANS flow simulation. As illustrated in Fig. \ref{fig:ForcedOscillationCurve}, the forced oscillation motion produces a hysteresis phenomenon on the resulting pitching moment coefficient curve.

According to \cite{Bhagwandin2014}, this phenomenon is almost constant and symmetrical around the user-chosen average angle $\alpha_0$ for most of the projectiles. In this case, the $(C_{mq}+C_{m\dot{\alpha}})_{\alpha=\alpha_0}$ coefficient can be calculated by using two points on the curve describing the evolution of the pitching moment coefficient during the oscillations. Both points are located on the $\alpha=\alpha_0$ axis and are represented by the $C_{m+}$ and $C_{m-}$ points corresponding to the Cm coefficients on the increasing and decreasing parts of the angle of attack, respectively. The $(C_{mq}+C_{m\dot{\alpha}})_{\alpha=\alpha_0}$ as well as \Cma coefficients can be calculated using following formulas \cite{Bhagwandin2014}:
\begin{equation}
	\left( C_{mq} + C_{m\dot{\alpha}} \right)_{\alpha=\alpha_0} = \frac{C_{m+} - C_{m-}}{2 k A} \text{\, for \,} k=\frac{\omega D}{2 V_\infty}
	\label{eq:ForcedCmqCma}
\end{equation}
\begin{equation}
	C_{m\alpha} = \frac{ \left( C_m \right)_{\alpha = \alpha_{max}} - \left( C_m \right)_{\alpha = \alpha_{min}}}{2 A}
	\label{eq:ForcedCma}
\end{equation}

\subsubsection{1Dof/CFD coupling}

This methodology consists in coupling CFD and rigid body dynamics (RBD) simulations so as to obtain a damped oscillating motion. As the simulations are performed exclusively along the pitch axis, the 1Dof denomination is used instead of RBD. The resulting $\alpha (t)$ signal is analysed with the MiRo post-processing algorithm, as described in part \ref{part:DeterminationCoefficients}. This methodology has the advantage of being very close to the free flight but, for the same reason as for the MiRo device, it can only be applied on stable configurations.

\subsubsection{Grid and time step convergence}

\begin{figure}[h]
	\graphicsvspace
	\centering
	\includegraphics[width=\linewidth]{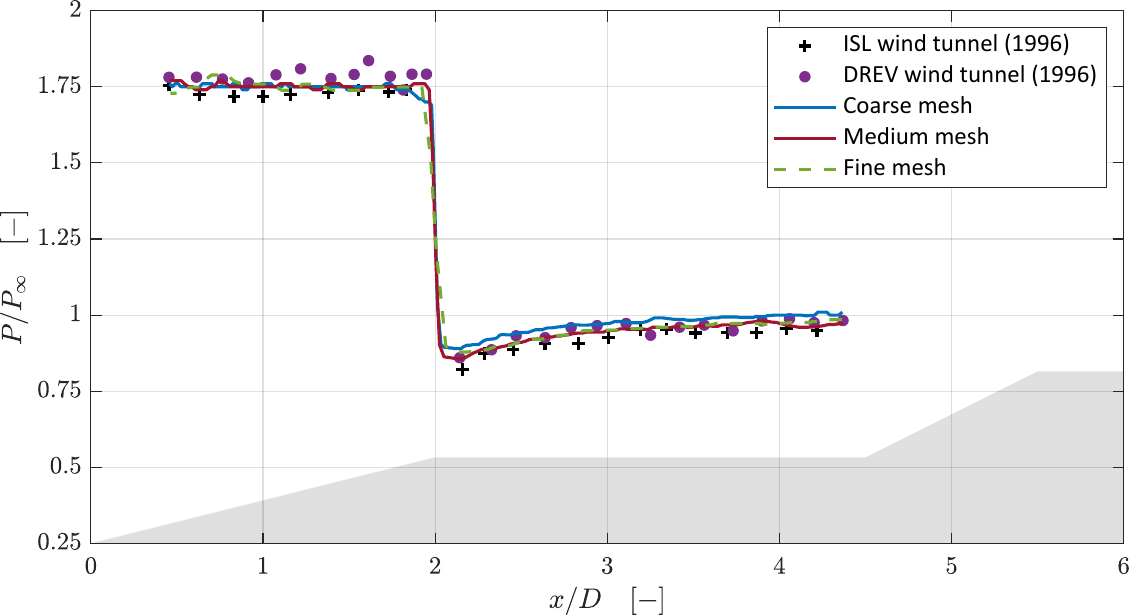}
	\captionvspace
	\caption{Pressure distribution along the longitudinal axis of the DREV-ISL projectile at Mach 2 and $\alpha$ = 5$^{\circ}$}
	\label{fig:distribution_de_pression_AOA_5_Mach_2}
	\figvspace
\end{figure}

A mesh independence study has been performed using RANS simulations in order to verify the influence of the spatial resolution on the pressure distribution of the projectile. Pressure distributions along the longitudinal axis were extracted for three fine (14.8 million cells), medium (5.6 million cells) and coarse (0.9 million cells) meshes. It was also checked that the $y^+ < 1$ criterion is being met on the entire geometry. Fig. \ref{fig:distribution_de_pression_AOA_5_Mach_2} compares the CFD results with the measurement obtained during a wind tunnel test campaign performed by Berner and Dupuis \cite{Berner1993} at Mach 2 and an angle of attack of 5$^{\circ}$. While the number of cells almost tripled, the pressure differences between the medium and fine meshes is invisible. Moreover, the pressure distributions obtained by CFD overlap the experimental data. For this reason, the intermediate mesh has been retained for the spatial discretization.

Independent time step convergence studies have been using both CFD calculation techniques. As the global time step $\Delta t$ depends on the Mach number, this study has been conducted in a Mach number range from 1.5 to 4.0 using velocity dependant relationships. Five inner iterations amounts have been investigated: $i = \left\{ 5, \, 10, \, 15, \, 20, \, 25 \right\}$.

For the forced oscillation motion technique, the reduced frequency has been set to $k = 0.1$ as suggested in \cite{Bhagwandin2014} and the amplitude to $A = 0.25^\circ$. Thus, $\Delta t$ has been calculated so as to decompose an oscillation period into $N = \left\{ 100, \, 200, \, 300, \, 400, \, 500 \right\}$ time steps. For each simulation, the $(C_{mq}+C_{m\dot{\alpha}})_{\alpha=\alpha_0}$ and the \Cma coefficients have been calculated using Eqs. \ref{eq:ForcedCmqCma} and \ref{eq:ForcedCma}. The velocity-independent convergence has been obtained for a  $\Delta t$ value corresponding to $N = 200$ and $i = 20$.

For the 1DoF/CFD coupling technique, a flow particle travelling an unknown ratio $r$ of the length of the projectile $L_{proj}$ has been considered, such as  $\Delta t = r \cdot L_{proj} / V_\infty$. The time step convergence has been fulfilled with $r = \frac{1}{14}$ and $i = 20$. For instance, these parameters lead to an amount of 1200 time steps per oscillation at Mach 2.0.

\section{COMPARISON OF THE RESULTS}

The stability derivative \Cma and the pitch damping coefficient sum \CmqCma are used to quantify the static- and dynamic stabilities of the ammunition, respectively. Using the body-fixed coordinate system defined in the DIN9300 norm \cite{din9300teil1}, decreasing values of these coefficients correspond to an improvement in terms of stability. A statically and dynamically stable projectile ($Cm_\alpha < 0$ and $C_{mq}+C_{m\dot{\alpha}} < 0$) tends to return to its balance angle of attack in case of disturbances during the flight. Static balance measurements facilitate the determination of the \Cma coefficient in wind tunnels \cite{libsig2018accuracy}. However, as the \CmqCma coefficient sum can only be obtained by means of moving experiments, the related one standard deviation ($\pm \upsigma$) percent error is commonly expected to be larger than 25\% \cite{prodasv3technical}.

\subsection{Discussion on the aerodynamic results}

\begin{figure}[h]
	\graphicsvspace
	\centering
	\includegraphics[width=\linewidth]{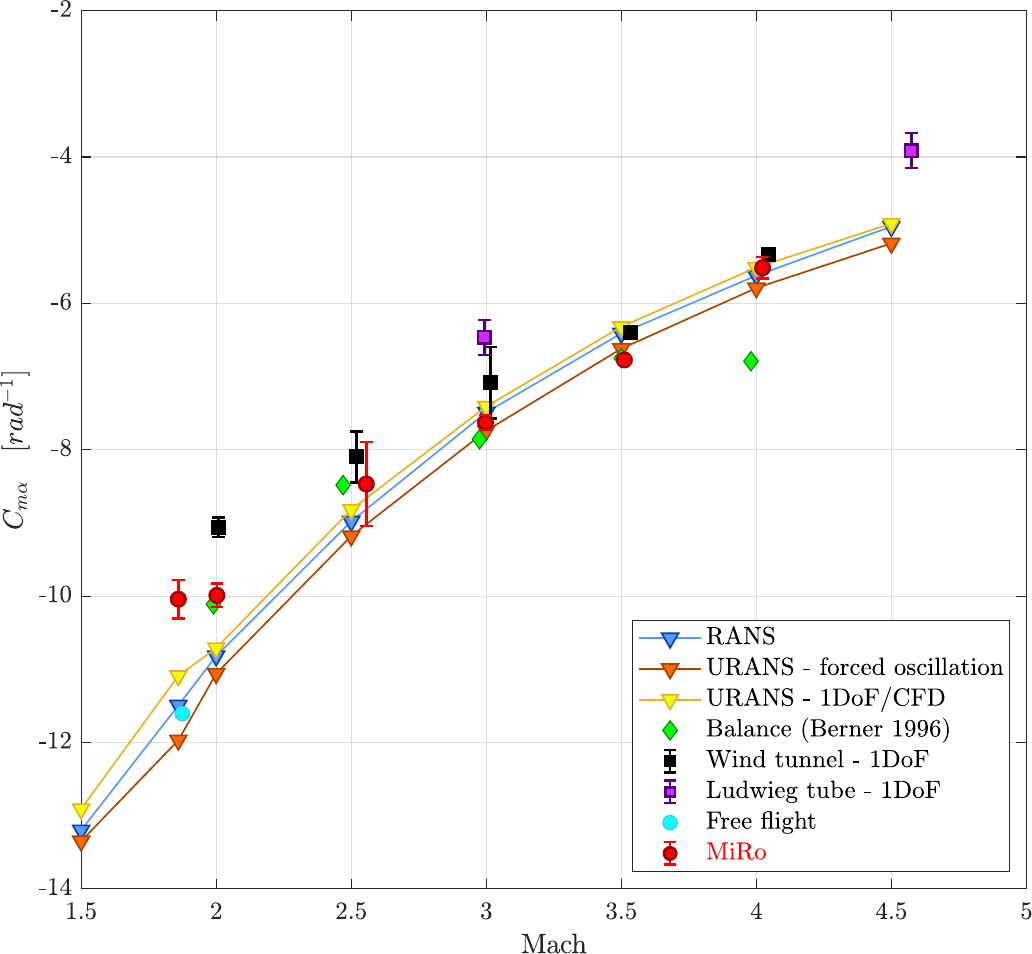}
	\captionvspace
	\caption{Comparison of the \Cma coefficients}
	\label{fig:comparaison_Cma_exp_CFD}
	\figvspace
\end{figure}

\begin{figure}[t]
	\centering
	\includegraphics[width=\linewidth]{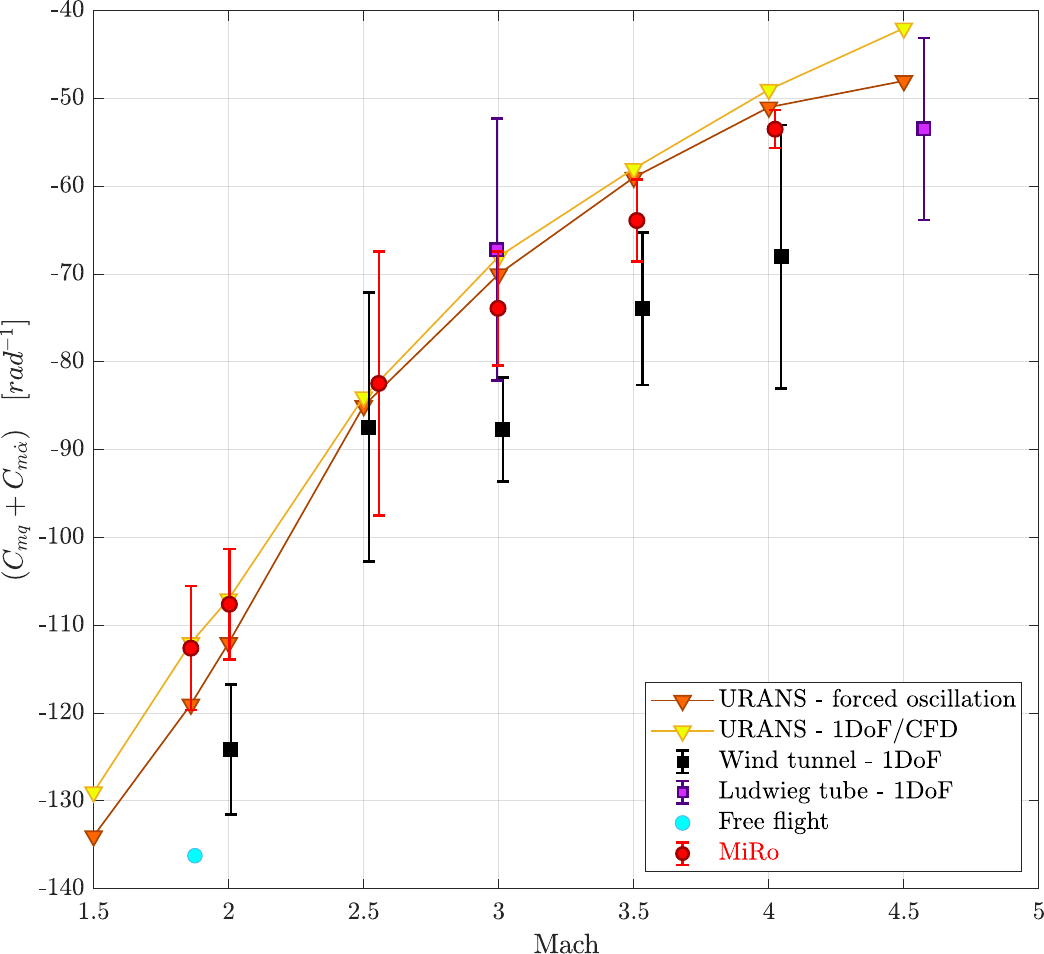}
	\captionvspace
	\caption{Comparison of the \CmqCma coefficients}
	\label{fig:comparaison_Cmq_exp_CFD}
\end{figure}

Figs. \ref{fig:comparaison_Cma_exp_CFD} and \ref{fig:comparaison_Cmq_exp_CFD} show the comparison of the \Cma and \CmqCma coefficients using the six methodologies described in part \ref{part:DescriptionFull}. Regarding the general shape of the curves, both \Cma and \CmqCma coefficients increase with the Mach number, meaning that the stability of the projectile decreases with an increasing velocity. This result is expected since the stabilizing tail-fins become more efficient during the deceleration of the supersonic projectile.

The MiRo measurements correspond to the red markers and their related error bars for a plus/minus two standard deviation ($\pm 2 \upsigma$). From a statistical point of view, each flow condition of the MiRo wind tunnel tests has been repeated ten times. The resulting MiRo dataset shows a one standard deviation ($\pm \upsigma$) percent error of 2 to 10\% depending on the Mach number, which is at least 2.5 times better than the commonly expected values.

The forced oscillation motion and 1DoF/CFD coupling results are displayed in orange and yellow, respectively. The MiRo measurements and the CFD results almost overlap each other. As the simulations have been performed for low angles of attack, namely an amplitude of 0.25$^{\circ}$ for the forced oscillations and an initial angle of attack of 2$^{\circ}$ for the 1DoF/CFD coupling, the accuracy of the CFD results is expected to be quite reliable. This observation provides additional confidence in the MiRo measurements.

Concerning the curves of Figs. \ref{fig:comparaison_Cma_exp_CFD} and \ref{fig:comparaison_Cmq_exp_CFD}, an excellent overall agreement has been obtained for all measurement and calculation techniques, qualitatively and quantitatively speaking. From a global point of view, the 1DoF freely rotating model held by a wire, in black and purple, provides lower \CmqCma than the MiRo measurements and the CFD calculations. Physically speaking, this trend can directly be explained by considering the interaction induced by the wire suspension. Indeed, as the wire suspension generates a wedge-shaped shock emanating from the centre of gravity of the model (location of the wire suspension), the Mach number at the rear of the projectile is lower than expected without suspension. In that case, as the global trend of Fig. \ref{fig:comparaison_Cmq_exp_CFD} indicates, the pitch damping coefficient sum of the wire-suspended techniques are expected to be lower than for a non-perturbed flight. This explains why the \CmqCma coefficients obtained when using the wire-suspended measurement technique are closer to the MiRo measurements if parasitic aerodynamic structures can be avoided.

The largest disagreement is obtained with the free flight-based aerodynamic coefficient determination technique. Nevertheless, even if free-flight corresponds to the most relevant technique, the precision of the MiRo measurements is hard to assess for the following reasons:
\itemizevspace
\begin{itemize}
\itemsep-.3em 
\item[-] Only four projectiles could be launched,
\item[-] Only non-instrumented ammunition could be launched for manufacturing and post-processing time-related reasons,
\item[-] As the projectiles decelerate during the flight, the flight conditions are not constant,
\item[-] The images of the trajectory trackers - the depicted projectiles being very small and the direction of the optical axis changing during flight - have been used for the stereovision-based post processing.
\end{itemize}
\itemizevspace
However, as the relative difference between the free-flight and the sums of the MiRo mean pitch damping coefficients sums amount to 17\%, the main conclusion to be drawn from these free-flight measurement is that the MiRo experiment provides consistent aerodynamic coefficients values. 

\subsection{Discussion on the effect of the aerodynamic coefficients differences on the shape of the trajectory}

Figs. \ref{fig:comparaison_Cma_exp_CFD} and \ref{fig:comparaison_Cmq_exp_CFD} show that the differences between the aerodynamic coefficients resulting from the six investigated techniques are relatively small. In order to quantify the effect of these differences on the shape of the real trajectory, the time evolution of the angle of attack modelled by Eqs. \ref{eq:ModeleAoA} - \ref{eq:ModeleAoA_B} is analysed. This analysis allows to calculate the following trajectory shaping variables (also called shape variables later on):
\itemizevspace
\begin{itemize}
\itemsep-.3em 
\item[-] $t_{50\%}$: time for which $\alpha (t_{50\%}) = 0.5 \alpha_{max}$, corresponding to the half-life period of the damped oscillation 
\item[-] $t_{95\%}$: time for which $\alpha (t_{95\%}) = 0.05 \alpha_{max}$, also considered to be the end of the damped oscillation
\item[-] $N_{95\%}$: number of oscillations performed by the projectile from $t_0$ to $t_{95\%}$
\item[-] $X_{95\%}$: distance travelled by the projectile from $t_0$ to $t_{95\%}$
\end{itemize}
\itemizevspace
The trajectories are evaluated by using the following constants: ISA atmospheric conditions at sea level, $D = 40 mm$, $Iy = 1.151 \times 10^{-3} kg.m^2$ and $\alpha_{max} = 5^{\circ}$.

\subsubsection{Effect of the two standard deviations errors}

\label{Part:TwoSigmaEffect}
This subsection focuses on the effect of the uncertainty of the MiRo measurements on the shape of the resulting trajectory. Thus, the quantities listed above have been calculated by using the worst MiRo measurement, namely the one obtained for Mach 2.5 (measurement with the largest \Cma and \CmqCma error bars). These results are listed in Tab. \ref{Tab:ErrorBarsComparisonMiRo} using the mean values and the extremities of the error bars (mean values $\pm 2 \upsigma$) of both investigated aerodynamic coefficients.

\begin{table}[t]
	\centering
	\begin{tabular}{|c|c|c|c|c|c|}
	\hline
	\CmaNoSpace & \hspace{-2.2pt} \CmqCmaNoSpace \hspace{-2.2pt} & $t_{50\%}$ & $t_{95\%}$ & $X_{95\%}$ & $N_{95\%}$ \\ \hline
	-7.89 & -67.4 & 44 ms & 0.19 s & 166 m & 12.1 \\ \hline
	-8.47 & -67.4 & 44 ms & 0.19 s & 166 m & 12.6 \\ \hline
	-9.05 & -67.4 & 44 ms & 0.19 s & 166 m & 13.0 \\ \hline
	-7.89 & -82.5 & 36 ms & 0.15 s & 136 m & 9.9  \\ \hline
	-8.47 & -82.5 & 36 ms & 0.15 s & 136 m & 10.3 \\ \hline
	-9.05 & -82.5 & 36 ms & 0.15 s & 136 m & 10.6 \\ \hline
	-7.89 & -97.5 & 31 ms & 0.13 s & 115 m & 8.4  \\ \hline
	-8.47 & -97.5 & 31 ms & 0.13 s & 115 m & 8.7  \\ \hline
	-9.05 & -97.5 & 31 ms & 0.13 s & 115 m & 9.0  \\ \hline
	\end{tabular}
	\tabcaptionvspace
	\caption{Effect of the MiRo error bars on the ballistic trajectory at Mach 2.5 for $\alpha_{max} = 5^{\circ}$}
	\label{Tab:ErrorBarsComparisonMiRo}
    \figvspace
\end{table}

As the \Cma is independent of the \CmqCma coefficient and directly linked to the oscillations frequency, only the $N_{95\%}$ variable can be impacted by a \Cma uncertainty. However, with respect to the results of Tab. \ref{Tab:ErrorBarsComparisonMiRo}, it can be assumed that the \Cma error bars of the MiRo measurements (red curve of Fig. \ref{fig:comparaison_Cma_exp_CFD}) are too small to have a noticeable effect on the real trajectory. 
On the other hand, the relative differences between the extremities of the \CmqCma error bar at Mach 2.5 are much bigger, leading to a non-negligible effect on the shape of the trajectory. The $X_{95\%}$ and $N_{95\%}$ variables are particularly impacted. Nevertheless, these conclusions have to be relativized, first, because the comparison has been based on the worst case, and second, because an exponential decay of 95\% takes a long time to be reached. This leads to big differences with respect to the threshold time. Globally, even if there are uncertainties for the MiRo \CmqCma measurements, the values of the trajectory shape variables remain consistent with each other.

\subsubsection{Effect of the coefficient differences resulting from the six coefficient determination techniques}

The full dataset of Fig. \ref{fig:comparaison_Cma_exp_CFD} shows that the largest differences between \Cma mean values are in the same order of magnitude as the length of the $\pm 2 \upsigma$ error bar obtained with the MiRo measurement at Mach 2.5. For this reason, as part \ref{Part:TwoSigmaEffect} identified the impact of the \Cma uncertainty to be negligible in comparison to the one of the \CmqCma, this analysis is only performed on the basis of the \CmqCma results of Fig. \ref{fig:comparaison_Cmq_exp_CFD}. The trajectory shape variables are calculated using the mean values exclusively. For the illustration, the shape variables are calculated for two flight points containing a large dataset, namely Mach 1.86 and Mach 3.0. The results are summarized in Tabs. \ref{Tab:MethodsComparisonMach186} and \ref{Tab:MethodsComparisonMach300}, respectively.

As expected (cf. results of Fig. \ref{fig:comparaison_Cmq_exp_CFD}), the largest discrepancies are obtained for the free flight at Mach 1.86 and the wire suspension at Mach 3.0. The other techniques, and especially the MiRo measurements, provide very consistent trajectory shape variables. This Mach number independent coherence builds additional trust in the MiRo measurements and shows that this methodology is very effective for pre-design or aerodynamic characterization studies.

The aerodynamic coefficients (Figs. \ref{fig:comparaison_Cma_exp_CFD} and \ref{fig:comparaison_Cmq_exp_CFD}) and trajectory shape variables (Tabs. \ref{Tab:MethodsComparisonMach186} and \ref{Tab:MethodsComparisonMach300}) obtained with the MiRo and the CFD determination techniques are very close to each other. This makes it all the more interesting because this consistence shows that the MiRo test bench could not only be used for validations of future CFD results but also as an experimental basis for comparison of transitory motions occurring in 6DoF trajectory simulations. Today, this technique is mature enough to be used for fast, low-cost and efficient pitch damping characterizations of fin-stabilized ammunition. 

\begin{table}[h]
	\centering
	\begin{tabular}{|c|c|c|c|}
	\hline
	Method & $t_{95\%}$ & $X_{95\%}$ & $N_{95\%}$ \\ \hline
	MiRo (Wind Tunnel) & 0.16 s	& 99 m & 8.2 \\ \hline
	Free flight & 0.13 s & 82 m & 7.3 \\ \hline
	Forced oscillation (CFD) & 0.15 s & 94 m & 8.5 \\ \hline
	1DoF/CFD coupling & 0.16 s & 100 m & 8.7  \\ \hline
	\end{tabular}
	\tabcaptionvspace
	\caption{Effect of the coefficient determination method on the ballistic trajectory at Mach 1.86 for $\alpha_{max} = 5^{\circ}$}
	\label{Tab:MethodsComparisonMach186}
    \figvspace
\end{table}

\begin{table}[h]
	\centering
	\begin{tabular}{|c|c|c|c|}
	\hline
	Method & $t_{95\%}$ & $X_{95\%}$ & $N_{95\%}$ \\ \hline
	MiRo (Wind Tunnel) & 0.15 s & 152 m & 10.9 \\ \hline
	Wire susp. (Wind Tunnel) & 0.12 s & 128 m & 8.9 \\ \hline
	Wire susp. (Ludwieg tube) & 0.16 s & 167 m & 11.0 \\ \hline
	Forced oscillation (CFD) & 0.16 s & 160 m & 11.6 \\ \hline
	1DoF/CFD coupling & 0.16 s & 165 m & 11.7 \\ \hline
	\end{tabular}
	\tabcaptionvspace
	\caption{Effect of the coefficient determination method on the ballistic trajectory at Mach 3.00 for $\alpha_{max} = 5^{\circ}$}
	\label{Tab:MethodsComparisonMach300}
    \figvspace
\end{table}

\section{CONCLUSION}

In this paper, MiRo, a novel and almost non-intrusive technique for dynamic wind tunnel measurements, has been evaluated by comparison with five other experimental and numerical methodologies. Despite the complex determination of the related physical values, very promising results have been obtained thanks to reliable measurements and low uncertainties.

This technique is based on the analysis of the time evolution of the attitude of a freely rotating model in the wind tunnel. As the final purpose of this technique consists in measuring the attitude in the 3 directions of space, stereovision technique has been employed for its development. However, if the amount of degrees of freedom is reduced to one or two, only one or two direct visualizations are necessary. This characteristics is very interesting for experiments with limited optical access.

Today, this technique has been successfully validated on a single degree of freedom. The next steps consist in progressively increasing its complexity by releasing the yaw and roll degrees of freedom, embedding an internal engine for the analysis decoupled ammunition, and in adding the possibility to investigate spin-stabilized projectiles.

\section{ACKNOWLEDGEMENTS}

The authors would like to thank all those who invested their time and resources in the development of the MiRo measurement technique and contributed to prepare and conduct the experiments. A special mention goes to the staff of the ISL's main workshop, the wind tunnel testing team, the shock tunnel testing team and the free flight testing team.

\bibliographystyle{plain}
\bibliography{references}

\end{document}